\documentclass[prd,jmp,reprint,onecolumn,superscriptaddress,12pt]{revtex4-2}

\usepackage{amssymb}
\usepackage{amsmath}
\usepackage{graphicx}
\usepackage{dcolumn}
\usepackage{bm}
\usepackage{color}
\usepackage{xcolor}
\usepackage{float}
\allowdisplaybreaks
\usepackage{subfig}
\usepackage{comment}
\usepackage{multirow}
\usepackage[figurename=FIGURE,tablename=TABLE]{caption}
\usepackage{threeparttable}
\usepackage{hyperref}
\hypersetup{
    colorlinks=true, 
    linktoc=all,     
    linkcolor=magenta,
    citecolor=red
}

\newcommand{\ba}{\begin{eqnarray}}
\newcommand{\ea}{\end{eqnarray}}

\begin{document}

\title{Thermalon mediated AdS to dS phase transitions in Einstein-Gauss-Bonnet-massive gravity}

\author{Supakchai Ponglertsakul}
\email{supakchai.p@gmail.com}
\affiliation{Strong Gravity Group, Department of Physics, Faculty of Science, Silpakorn University, Nakhon Pathom 73000, Thailand}

\author{Phongpichit Channuie}
\email{channuie@gmail.com}
\affiliation{College of Graduate Studies, Walailak University, Thasala, Nakhon Si Thammarat, 80160, Thailand}
\affiliation{School of Science, Walailak University, Thasala, Nakhon Si Thammarat, 80160, Thailand}

\author{Daris Samart}
\email{darisa@kku.ac.th}
\affiliation{Khon Kaen Particle Physics and Cosmology Theory Group (KKPaCT), Department of Physics, Faculty of Science, Khon Kaen University, 123 Mitraphap Rd., Khon Kaen, 40002, Thailand}


\date{\today}

\begin{abstract}

In this work, gravitational phase transition emerging from anti de-Sitter (AdS) to de-Sitter (dS) vacua in Einstein-Gauss-Bonnet-massive gravity (EGBMG) is considered. We determine the location of thermalon (a static bubble solution in Euclidean space) which exists in casually connected two regions of the spacetime. The thermalon plays a major role in gravitational phase transition by inducing the decay of the negative effective cosmological constant to the positive one due to the higher-order gravity effects. The thermodynamics phase space of the Hawking temperature and free energy is investigated in details. We find that the free energy of the thermalon is always negative at the maximum of Hawking temperature for all possible values of the Gauss-Bonnet couplings. 
This means that the phase transition mediated by thermalon from AdS to dS asymptotics is inevitable 
according to the presence of the massive gravity. More importantly, the parameters of the massive gravity behave similarly to the order parameters in the phase transition instead of the Gauss-Bonnet coupling. 
\end{abstract}

\keywords{Bubble nucleation, Higher dimensional gravity, Gauss-Bonnet gravity, Massive gravity, AdS/dS gravitational phase transition }
\maketitle

\newpage
\section{Introduction}

The Cosmological Constant (CC) problem has been a major challenge in field theory and cosmology for several decades. On the one hand, the CC is interpreted as the vacuum energy density and the prediction of the standard quantum field theory calculation gives the theoretical result up to the Planck scale bigger than the observational value about 120 order of magnitudes \cite{Weinberg:1988cp}. This contradiction poses a serious problem and remains as the unsolved problem in high energy physics. On the other hand, the CC can be used to describe the accelerating expansion of the universe at the present time or the dark energy problem \cite{Peebles:2002gy,Padmanabhan:2002ji,Carroll:2000fy}. Nowadays, the interpretation of the CC as dark energy becomes the standard model of cosmology known as the Lambda Cold Dark Matter (LCDM) and the LCDM model can describe most of the cosmological data. However, according to the CC problem, we need the correct theory or better solution for explaining the fundamental mechanism of the CC problem. This leads to a number of the theoretical models proposed to solve the smallness of observed values of the CC. In particular, an approach from the Beyond Standard Model (BSM) of particle physics is received a lot of attention, for instances, supersymmetric model, large extra dimension approach of CC, string landscape and etc. See Refs.\cite{Burgess:2004kd,Banks:2014pqa,Mavromatos:2015jzo,Burgess:2007ui,Lue:2005ya,Bull:2015stt,Polchinski:2006gy,Denef:2007pq,Douglas:2019kus} for reviews on this subject.   
One of interesting approach for describing the smallness of the CC value is the gravitational phase transition. The phase transition arises from the bubble nucleation of the true vacuum in the false vacuum spacetime when the bubble pops up and expands until filling up the whole spacetime. This means that the CC of the false vacuum is changed to the new value of the CC according to new vacuum energy of the true vacuum. The mechanism of this gravitational phase transition is originally proposed by Refs.\cite{Coleman:1977py,Coleman:1980aw}. The phase transition for Einstein gravity coupled to the 3-form fields is studied and the results revealed that the CC can be dynamically changed into several values from the infinite decay of the false vacuum and this leads to an alternative approach to describe the CC problem \cite{Brown:1987dd,Brown:1988kg}. More interestingly, Ref.\cite{Gomberoff:2003zh} has used the same gravitational setting but hosting the black hole inside the bubble. Then a thermalon, the instanton solution of the thermal trigger for the bubble nucleation, plays the role as a mediator of the phase transition. Under this consideration, the bubble with the three-form charge induces the decay of the dS false vacuum into another smaller dS true vacuum. In order to thermally trigger the phase transition, one needs the source of the exotic matter fields with negative pressure. Alternatively, Refs.\cite{Camanho:2012da,Camanho:2013uda} have proposed new mechanism analogous to the thermalon mediated the phase transition by introducing the higher order gravity into the system. There are two solution branches of the higher order gravity corresponding to negatively and positively effective cosmological constants (AdS and dS spacetime, respectively). 
Initially, the false vacuum is displayed by the AdS spacetime. The true vaccum is represented by the dS geometry emerged from the creation of bubble with black hole inside. The bubble then expands very rapidly until it reaches the cosmological horizon. This changes the values of the effective CC from negative to positive values due to the thermalon mediated AdS to dS phase transition without inclusion of any matter fields coupled to the gravity. This is the salient feature of this type of gravitational phase transition.

On the other hand, the existence of the graviton mass provides an interesting results in either quantum theory of gravity and cosmology (see \cite{Hinterbichler:2011tt,deRham:2014zqa} for reviews). A consideration of the graviton mass has been received attentions for several decades \cite{deRham:2016nuf}. In particular, a so-called dRGT massive gravity \cite{deRham:2010kj} becomes a well defined theory of massive gravity. Several problems of the graviton mass is solved by using the dRGT massive gravity approach, for instances, redundancy of the spin d.o.f. \cite{Fierz:1939ix}, vDVZ discontinuous \cite{vanDam:1970vg,Zakharov:1970cc}, Vainshtein screening \cite{Vainshtein:1972sx} and the unstable ghost modes of the massive gravity \cite{Boulware:1972yco}. In addition, massive gravity and its extentions can be used to describe dark energy \cite{DAmico:2011eto,Akrami:2012vf,Akrami:2015qga}. Moreover, the dRGT massive gravity is also extensively studied in various aspects, for examples, black hole solution and its thermodynamics properties \cite{Ghosh:2015cva}, quasinormal modes \cite{Burikham:2017gdm}, dark matter \cite{Panpanich:2018cxo}, gravitational lensing \cite{Panpanich:2019mll} and wormholes \cite{Tangphati:2020mir}. 

In this work, we will utilize the thermalon formalism with the higher order gravity to study the AdS to dS phase transition by including the effect of the massive gravity in five dimensional spacetime. The main motivation of this work is that the Gauss-Bonnet term is non-trivial in the spacetime dimension higher than four \cite{Lovelock:1971yv} and it has been widely used in brane cosmology \cite{Charmousis:2002rc, Sheykhi:2007gi, Deruelle:2000ge, Cai:2005ie, Lidsey:2003sj, Lidsey:2002zw}. Interestingly, one can consider the effects of the graviton mass since the massive graviton is well defined in the large extra dimension of the brane cosmology paradigm \cite{Gabadadze:2003jq,Cho:2001su}. Therefore, it is worth exploring the gravitational phase transition with the massive gravity in five dimensional spacetime. Charged black holes in Gauss-Bonnet-massive gravity and their thermodynamic properties is explored in \cite{Hendi:2015pda}. Phase transition of charged black hole in Gauss-Bonnet-massive gravity coupled with Yang-Mills field is investigated in \cite{Meng:2016its}. Moreover, the Gauss-Bonnet gravity in five dimension is employed to study the thermalon mediated AdS to dS phase transition in various aspects, e.g., study the critical point of the phase transition in the vacuum solution \cite{Camanho:2015zqa}, phase transition in black hole chemistry picture \cite{Hennigar:2015mco} and inclusions of Maxwell \cite{Samart:2020qya} and Kalb-Ramond \cite{Samart:2020mnn} fields.    

We will organize the paper as follows. The gravitational action of the higher order gravity with the massive gravity is set up and the corresponding branch solutions are given in section \ref{sec2}. The junction condition between two regions of AdS and dS vacua is derived and the dynamics of bubble nucleation is analyzed in section \ref{sec3}. In section \ref{sec4}, the gravitational phase transition in the thermodynamics phase space is studied in details. Finally, we close the paper with the conclusion in the section \ref{sec5}.  

\section{The Einstein-Gauss-Bonnet with the presence of the Massive gravity}\label{sec2}
In this section, we will set up the system of the AdS to dS gravitational phase transition via the thermalon mediation. The five dimensional action of Einstein-Gauss-Bonnet-massive gravity (EGBMG) with a presence of the bare CC ($\Lambda$) is given by
\begin{align}
    S &= \int d^5x \sqrt{-g}\left[R-2\Lambda+\alpha \mathcal{L}_{GB} + m^2 \mathcal{U}(g,\phi^a)\right], \label{action}
\end{align}
where $R$ is Ricci scalar, $\alpha$ is the Gauss-Bonnet coupling constant carrying the mass dimension $-2$ and $m$ can be interpreted as graviton mass. We have set $16\pi G_5 = 1$ with $G_5$ is Newton constant in five dimensions through out this work. The Gauss-Bonnet term takes a form
\begin{align}
    \mathcal{L}_{GB} &= R_{\mu\nu\rho\sigma}R^{\mu\nu\rho\sigma} - 4 R_{\mu\nu}R^{\mu\nu} + R^2.
\end{align}
The graviton potential $\mathcal{U}$ is a function of metric tensor $(g)$ and St\"{u}ckelberg field $(\phi^a)$. It is defined by
\begin{align}
    \mathcal{U} &= \sum_{i=1}^{4}c_i \mathcal{U}_i,
\end{align}
where $c_i$ are constants and
\begin{align}
    \mathcal{U}_1 &= \left[ \mathcal{K} \right], \nonumber \\
    \mathcal{U}_2 &= \left[ \mathcal{K} \right]^2 - \left[ \mathcal{K}^2 \right] , \nonumber \\
    \mathcal{U}_3 &= \left[ \mathcal{K} \right]^3 -3 \left[ \mathcal{K} \right]\left[ \mathcal{K}^2 \right] + 2 \left[ \mathcal{K}^3 \right], \nonumber \\
    \mathcal{U}_4 &= \left[ \mathcal{K} \right]^4 -6\left[ \mathcal{K}^2 \right]\left[ \mathcal{K} \right]^2 + 8\left[ \mathcal{K}^3 \right]\left[ \mathcal{K} \right] + 3\left[ \mathcal{K}^2 \right]^2 - 6 \left[ \mathcal{K}^4 \right].
\end{align}
The matrix $K^{\mu}_{\nu} = \delta^{\mu}_{\nu} - \sqrt{g^{\mu\sigma}f_{ab}\partial_{\sigma}\phi^a\partial_{\nu}\phi^b}$, $[\mathcal{K}]=\mathcal{K}^{\mu}_{\mu}$ and $[\mathcal{K}^n]=(\mathcal{K}^n)^{\mu}_{\mu}$, where $f_{ab}=diag(0,0,\frac{c^2}{1-k\rho^2},c^2\rho^2,c^2\rho^2\sin^2\theta)$ is called fiducial metric with constants $c$ and $k$.
Varying the action (\ref{action}) with respect to $g^{\mu\nu}$ yields Einstein field equations,
\begin{align}
    G_{\mu\nu} + \Lambda g_{\mu\nu} + \alpha T_{\mu\nu}^{GB} + m^2 T_{\mu\nu}^{MG} &= 0\,. \label{EFE}
\end{align}
The effective energy-momentum tensors, $T_{\mu\nu}^{GB}$ and $T_{\mu\nu}^{MG}$ are defined by
\begin{align}
        T_{\mu\nu}^{GB} &= 2\left[R_{\mu\nu}R-\nabla_\nu\nabla_\mu R\right] + 4 \left[\nabla^{\rho}\nabla^{\sigma} R_{\mu\rho\nu\sigma} + \nabla^{\sigma}\nabla^{\rho} R_{\mu\rho\nu\sigma} + R_{\mu\rho\sigma\delta}R_{\nu}^{\rho\sigma\delta}\right] \nonumber \\
    &~~~-\frac{1}{2}g_{\mu\nu}\left[R^2 + 2 R_{\rho\sigma\delta\lambda}R^{\rho\sigma\delta\lambda} -4\Box R \right], \nonumber \\
    T_{\mu\nu}^{MG} &= -\frac{c_1}{2}g_{\mu\nu}\left(\mathcal{U}_1-\mathcal{K}_{\mu\nu}\right)-\frac{c_2}{2}\left(g_{\mu\nu}\mathcal{U}_2 - 2\mathcal{U}_1\mathcal{K}_{\mu\nu}+2g_{\rho\nu}\mathcal{K}_{\sigma}^{\rho}\mathcal{K}_{\mu}^{\sigma}\right) \nonumber \\
    &-\frac{c_3}{2}\left(g_{\mu\nu}\mathcal{U}_3+3\mathcal{U}_2g_{\mu\nu} + 6g_{\rho\nu}\mathcal{U}_1 \mathcal{K}_{\sigma}^{\rho}\mathcal{K}_{\mu}^{\sigma} -6g_{\rho\nu}\mathcal{K}_{\sigma}^{\rho}\mathcal{K}_{\lambda}^{\sigma}\mathcal{K}_{\mu}^{\lambda}\right) \nonumber \\
    &-\frac{c_4}{2}\left(g_{\mu\nu}\mathcal{U}_4 - 4 \mathcal{U}_3\mathcal{K}_{\mu\nu} + 12g_{\rho\nu}\mathcal{U}_2\mathcal{K}_{\sigma}^{\rho}\mathcal{K}_{\mu}^{\sigma}  - 24g_{\rho\nu}\mathcal{U}_1 \mathcal{K}_{\sigma}^{\rho}\mathcal{K}_{\mu}^{\sigma} 
    + 24 g_{\rho\nu}\mathcal{K}_{\sigma}^{\rho}\mathcal{K}_{\lambda}^{\sigma}\mathcal{K}_{\alpha}^{\lambda}\mathcal{K}_{\mu}^{\alpha}\right).
\end{align}
The static spherically symmetric with constant of spatial curvature $k$ is given by
\begin{align}
ds^2 &= -f(r)dt^2 + f^{-1}dr^2 + r^2 \left(\frac{d\rho^2}{1-k\rho^2} + \rho^2d\theta^2 + \rho^2\sin^2\theta d\phi^2\right).
\label{line-element}
\end{align}
The solutions of the metric tensor, $f(r)$ in Eq.(\ref{line-element}) is obtained by solving Einstein field equations (\ref{EFE}). It reads
\begin{align}
f_{\pm} &= k + \frac{r^2}{4\alpha}\left(1 \pm \sqrt{1 + \frac{16k^2\alpha^2}{r^4} + \alpha\left(\frac{4\Lambda}{3} + \frac{16m_0}{r^4} \right) - m^2\alpha\left( \frac{8cc_1}{3r} + \frac{8c^2c_2}{r^2} + \frac{16c^3c_3}{r^3}\right)  } \right).   \label{metricfn}   
\end{align}
Here $m_0$ is integration constant which can be interpreted as black hole's mass. If we set graviton mass to zero, $m=0$, this solution can be compared with those found in \cite{Camanho:2015zqa,Hennigar:2015mco,Samart:2020qya,Samart:2020mnn}. Thus, it is more convenient to define the following quantities
\begin{align}
    \alpha \to \frac{\lambda L^2}{2},~~~\Lambda \to \frac{6}{L^2},~~~m_0 \to \frac{\mathcal{M} -\lambda k^2 L^2}{2},
\end{align}
where $L$ is the AdS or dS length scale and $\lambda$ is the dimensionless Gauss-Bonnet coupling. We also define shorted-hand notation for the massive graviton term as 
\begin{align}
    c c_1 m^2 \to \gamma_a,~~~ c^2 c_2 m^2 \to \gamma_b,~~~ c^3 c_3 m^2 \to \gamma_c.
\end{align}
Therefore, the metric describing AdS outer ($+$) and dS inner ($-$) spacetime are written by
\begin{align}
f_{\pm} &= k + \frac{r^2}{2\lambda L^2} \left( 1 \pm \sqrt{1 + 4\lambda \left(1 + L^2 \left[ \frac{\mathcal{M}}{r^4} - \frac{\gamma_a}{3r} - \frac{\gamma_b}{r^2} - \frac{2\gamma_c}{r^3}     \right]\right)} \right).
\label{metric-fpm}
\end{align}
Alternatively, the solution of EGBMG can be obtained by introducing the following polynomials function
\begin{align}
    \Upsilon[g] &= -\frac{1}{L^2} + g + \lambda L^2 g^2 = \frac{\mathcal{M}}{r^4} - \frac{\gamma_a}{3r} - \frac{\gamma_b}{r^2} - \frac{2\gamma_c}{r^3}. \label{Upsilonfunc}
\end{align}
Thus the function $g$ can be explicitly written as
\begin{align}
g_{\pm} &= - \frac{1}{2L^2\lambda}
\left[1 \pm \sqrt{1 + 4\lambda 
\left( 1+L^2 
\left( \frac{\mathcal{M}}{r^4} - \frac{\gamma_a}{3r} - \frac{\gamma_b}{r^2} - \frac{2\gamma_c}{r^3} 
\right)  \right)} \right], 
\end{align}
or equivalently,
\begin{align}
    g_{\pm} &= \frac{k-f_{\pm}}{r^2}.
\end{align}
We will use two branches of solution Eq.(\ref{metric-fpm}) of the EGBMG theory for studying the gravitational phase transition in the next section.

\section{Junction condition and bubble dynamics}\label{sec3}
In this section, we will define the junction condition connecting two region (AdS and dS as outside and inside, respectively) of the spacetimes to study the bubble dynamics. To ensure the continuity between the two spacetimes, the so-called junction conditions must be satisfied. The junction condition of EGBMG is defined in terms of co-moving time component of the canonical momenta ($\Pi$) (see detail derivation in Refs.\cite{Samart:2020qya,Camanho:2015zqa,Camanho:2013uda})
\begin{align}
\Pi_{\pm} &= \frac{2}{3}\lambda L^2 g_{\pm}\sqrt{H-g_{\pm}}+\frac{4}{3}H\lambda L^2 \sqrt{H-g_{\pm}} + \sqrt{H-g_{\pm}}, \label{eqg}
\end{align}
where $H=\frac{k+\Dot{a}^2}{a^2}$. Here $a\equiv a(\tau)$ is a scaled factor defined on the induced line elements of the timelike hypersurface ($\Sigma$). We remark that, at the boundaries outer AdS and inner dS spacetimes, the metric of hypersurface $\Sigma$ takes the similar form (see more details in \cite{Samart:2020qya}). The derivative with respect to co-moving time coordinate $\tau$ will be denoted by $\dot{a}=\frac{da}{d\tau}$.

The junction condition can be obtained from
\begin{align}
\Pi_{+}^2 - \Pi_{-}^2 \rvert_{r_{\pm}=a} &= 0,
\label{junction-con}
\end{align}
where this condition is evaluated at the boundaries between two manifolds (outer and inner spacetimes). By substituting Eq.(\ref{eqg}) into the above condition, we obtain 
\begin{align}
-\Dot{a}^2 &= k + \frac{a^6}{4L^2\lambda \mathcal{A}}\left(g_+\left(3+2g_+ L^2\lambda\right)^2 - g_-\left(3+2g_- L^2\lambda\right)^2\right), 
\end{align}
where
\begin{align}
\mathcal{A} &\equiv 3\left(\mathcal{M}_+-\mathcal{M}_-\right)  - a^3 \left(\gamma_{a+}-\gamma_{a-}\right)  -3a^2 \left(\gamma_{b+}-\gamma_{b-}\right) - 6a \left(\gamma_{c+}-\gamma_{c-}\right).    
\end{align}
The junction condition can be expressed in term of kinetic and effective potential as
\begin{align}
\frac{1}{2}\Dot{a}^2 + V(a) =0.
\end{align}
In order to obtain above equation, we have used the imaginary time, i.e., $\tau \to i\tau_E$ in the derivative leading to $(da/d\tau)^2 \to -(da/d\tau_E)^{2}$.
The effective potential can be shown explicitly
\begin{align}
V &= \frac{a^6}{8L^2\lambda \mathcal{A} }\left[4\left(2+g L^2\lambda\right)\left(\frac{\mathcal{M}}{a^4} - \frac{\gamma_a}{3a} - \frac{\gamma_b}{a^2} - \frac{2\gamma_c}{a^3}\right) + g(1+4\lambda)\right]_{-}^{+} + \frac{k}{2},
\label{effpot}
\end{align} 
where we have used the following notation
\begin{align}
\left[\mathcal{O}\right]_{-}^{+} &\equiv \mathcal{O}_+ - \mathcal{O}_-.
\end{align}
The derivative of effective potential with respect to $a$ is determined by
\begin{align}
V'(a) &= 
\frac{a}{8L^2\lambda\mathcal{A}^2}\Big(3a^5\Big[ a^2 \gamma_{a} + 2 a  \gamma_{b} + 2 \gamma_{c+}\Big]_-^+ \Big[g(1+4\lambda) + 4(2+gL^2\lambda)\Upsilon[g(a)]\Big]_{-}^{+}
\nonumber\\
&\qquad\qquad\quad + \Big[ 12\mathcal{M} + 6a^4 (1+4\lambda)g - \gamma_{a} a^3 (13+6gL^2\lambda) \nonumber\\
&\qquad\qquad\qquad\quad - 6\gamma_b a^2 (5+2gL^2\lambda) - 6\gamma_c a (7+2gL^2\lambda) \Big]_{-}^{+}
\Big),
\label{diff-effpot}
\end{align}
The effective potential in Eq.(\ref{effpot}) and its derivative in Eq.(\ref{diff-effpot}) are useful to explain the dynamics of the bubble nucleation and the discussion will be given later. More importantly, we shall assume the continuity of gravity mass by setting
\begin{align}
    \gamma_a = \gamma_{a_+} = \gamma_{a_-}, \quad
    \gamma_b = \gamma_{b_+} = \gamma_{b_-}, \quad
    \gamma_c = \gamma_{c_+} = \gamma_{c_-}.
\end{align}
These continuities greatly simplify the expression of effective potential. Imposing $V(a_{\ast})=V'(a_{\ast})=0$, one can explicitly determined $\mathcal{M}_{\pm}$ at the thermalon location, $a_{\ast}$ as
\begin{eqnarray}
\mathcal{M}^{\ast}_\pm &=& \frac{1}{12\lambda L^2 a_{\ast}^4}
\Big[ 3 a_{\ast}^6(1+4\lambda)\big(6k\lambda L^2 +  a_{\ast}^2(3+2g^{\ast}_\mp\lambda L^2)\big) 
\nonumber \\
&&  - 2\lambda L^2 a_{\ast}^5\big(9k\lambda L^2 + a_{\ast}^2(4 + 3g^{\ast}_\mp\lambda L^2)\big)\gamma_a 
\nonumber \\
&&  - 12\lambda L^2 a_{\ast}^4\big(3k\lambda L^2 + a_{\ast}^2(1 + g^{\ast}_\mp\lambda L^2)\big)\gamma_b 
-12 \lambda^2 L^4 a_{\ast}^3 (3k+a_{\ast}^2 g^{\ast}_\mp)\gamma_c \Big],
\label{thermalon-mass}
\end{eqnarray}
where $g^{\ast}_{\pm}=g_{\pm}(a_{\ast})$. From definition of the $\Upsilon$ in Eq.(\ref{Upsilonfunc}), we obtain
\begin{align}
    \Upsilon[g^{\ast}_\pm] 
    &= C_1 g^{\ast}_\pm + C_2,
\end{align}
where the coefficients $C_{1,2}$ are then determined as
\begin{align}
    C_1 &= \frac{1}{2} + \left[2 - \frac{L^2}{a_{\ast}}\left(  \frac{\gamma_a}{2} + \frac{\gamma_b}{a_{\ast}} + \frac{\gamma_c}{a_{\ast}^2}\right)\right]\lambda, \label{coeff1}\\
    C_2 &= \frac{3\left(1+4\lambda\right)}{4\lambda L^2} -  \frac{3k\left(1+4\lambda\right)}{2a_{\ast}^2}  + \frac{\gamma_a}{a_{\ast}} + \frac{2\gamma_b}{a_{\ast}^2} + \frac{2\gamma_c}{a_{\ast}^3}. \label{coeff2}
\end{align}
Thus we can solve these equations simultaneously which yield
\begin{align}
    g^{\ast}_\pm &= -\frac{\left(1+C_1\right) \mp \sqrt{1-2C_1-3C_1^2+4\lambda\left(1+C_2 L^2\right)}}{2\lambda L^2}. 
\end{align}
We note that the $g_\pm^\ast$ correspond to AdS and dS geometries and the condition $g_+^\ast \neq g_-^\ast$ is required since we study the phase transition between those spacetimes. Therefore by using Eq.(\ref{coeff1}--\ref{coeff2}), the $g^{\ast}_{\pm}$ can be expressed in terms of $k,\lambda,L^2,\gamma_a,\gamma_b$ and $\gamma_c$. After deriving all relevant equations, we are ready to analyze the bubble dynamics.
\begin{figure}[h]
    \centering
    \includegraphics[width=0.45\textwidth]{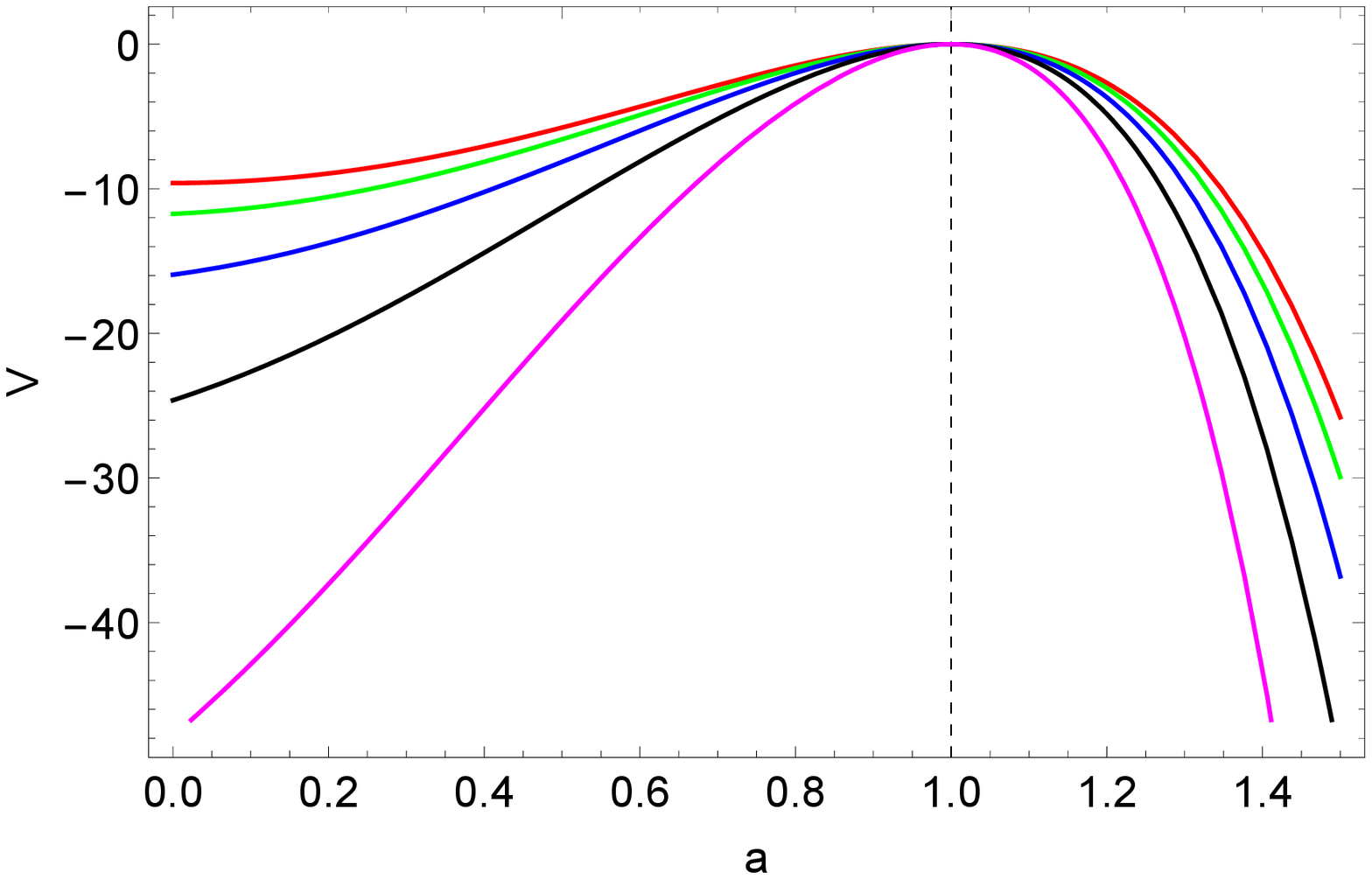}
    \includegraphics[width=0.45\textwidth]{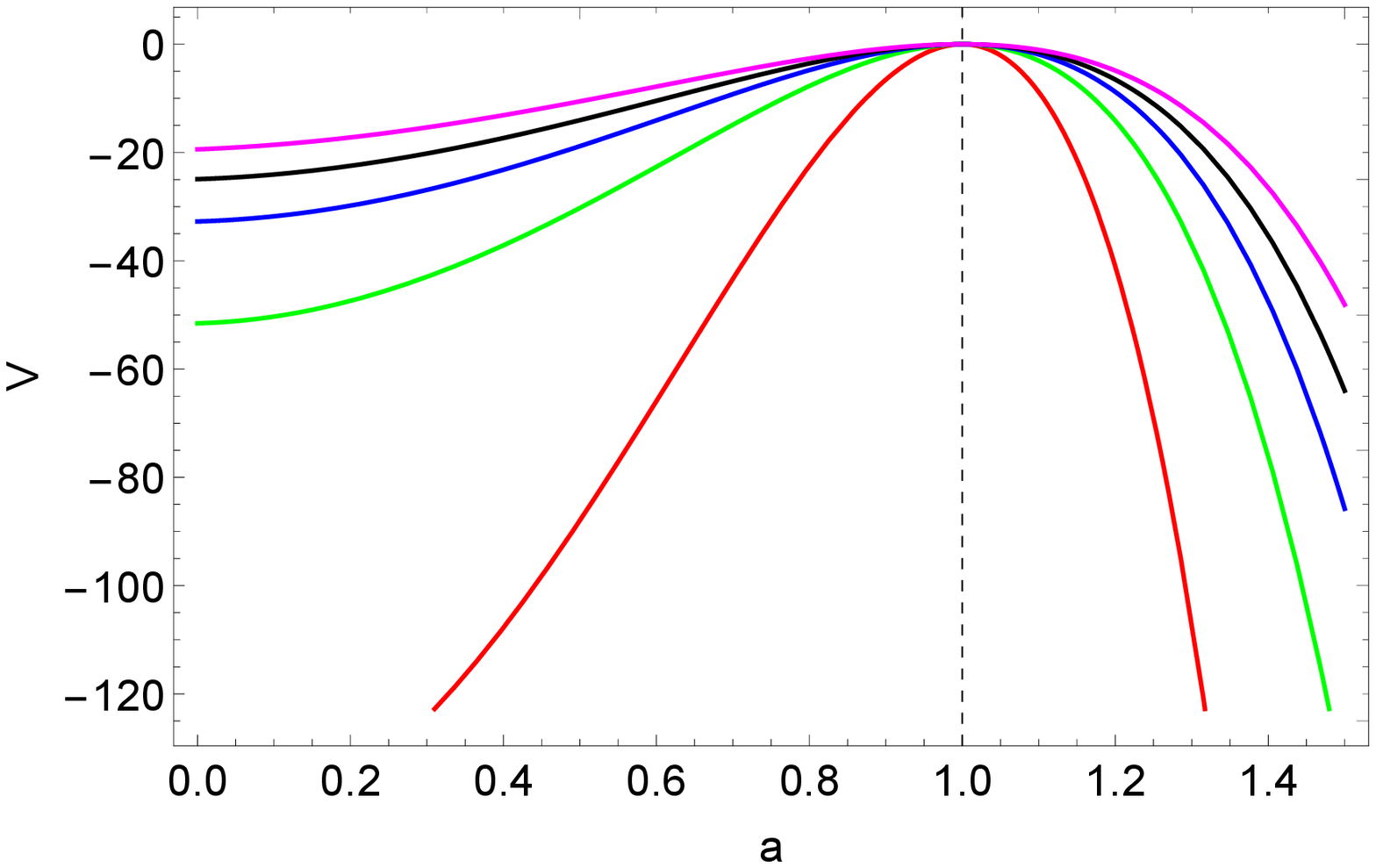}  
\caption{Effective potential plotted as function of $a$ for $a_{\ast}=1$, $L=1$, $k=1$: Left fixed $\lambda=0.015$ and $\gamma_i=$ 0, 2.5, 5, 7.5, 10 (red, green, blue, black, magenta), Right fixed $\gamma_i=5$ and $\lambda=$ 0.001, 0.003, 0.005, 0.007, 0.01 (red, green, blue, black, magenta), where $i=$ \{$a$, $b$, $c$\}
} \label{fig:figVeff}
\end{figure}
From now on, the spherical bubble, $k=1$ will be studied throughout this work. In Fig.~\ref{fig:figVeff}, we display an example of behaviour of effective potential against $a$. In this figure, we fix $\gamma_a=\gamma_b=\gamma_c$. We vary $\gamma_i$ while keep other parameters fixed in the left figure. The right figure shows the effective potential with $\lambda$ is varied. Despite having various parameters, throughout this work, we shall particularly focus on the set of parameter which yield $\mathcal{M}^*_{\pm}$ are real positive value. As the results, the Fig.~\ref{fig:figVeff} clearly shows that the effective potential has the maximum value at the given thermalon position, $a_\ast = 1$. This means that the thermalon is in the unstable mode and ready to roll down from the top of the effective potential. Consequently, the bubble, representing dS space with true vacuum can expand and leading to the AdS to dS phase transition eventually. In Fig.~\ref{fig:figMpara}, parameter spaces where $\mathcal{M}^*_{\pm}>0$ are illustrated. In each plots, we fix bubble location and AdS (dS) length scale to be unity. The GB coupling constant is chosen to be 0.1. With given $\gamma_c$, allowed value of $\gamma_a,\gamma_b$ can be obtained. The shaded regions increase as $\gamma_c$. We should emphasize that the plots in Fig.~\ref{fig:figMpara} are only partially displayed for demonstration purpose. From our numerical investigation, the shaded regions can be extended rightward and upward.

\begin{figure}[h]
    \centering
    \includegraphics[width=0.32\textwidth]{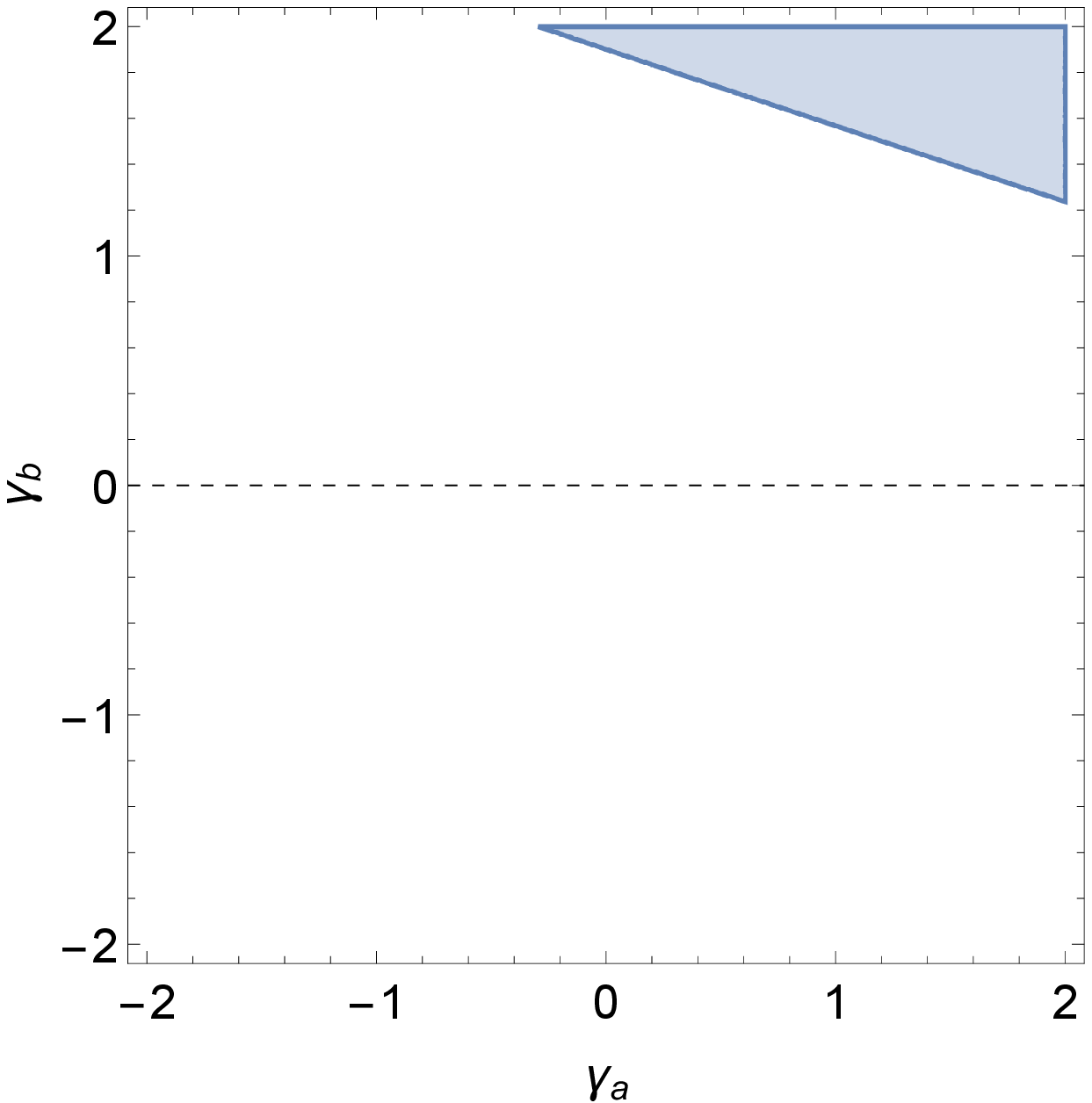}
    \includegraphics[width=0.32\textwidth]{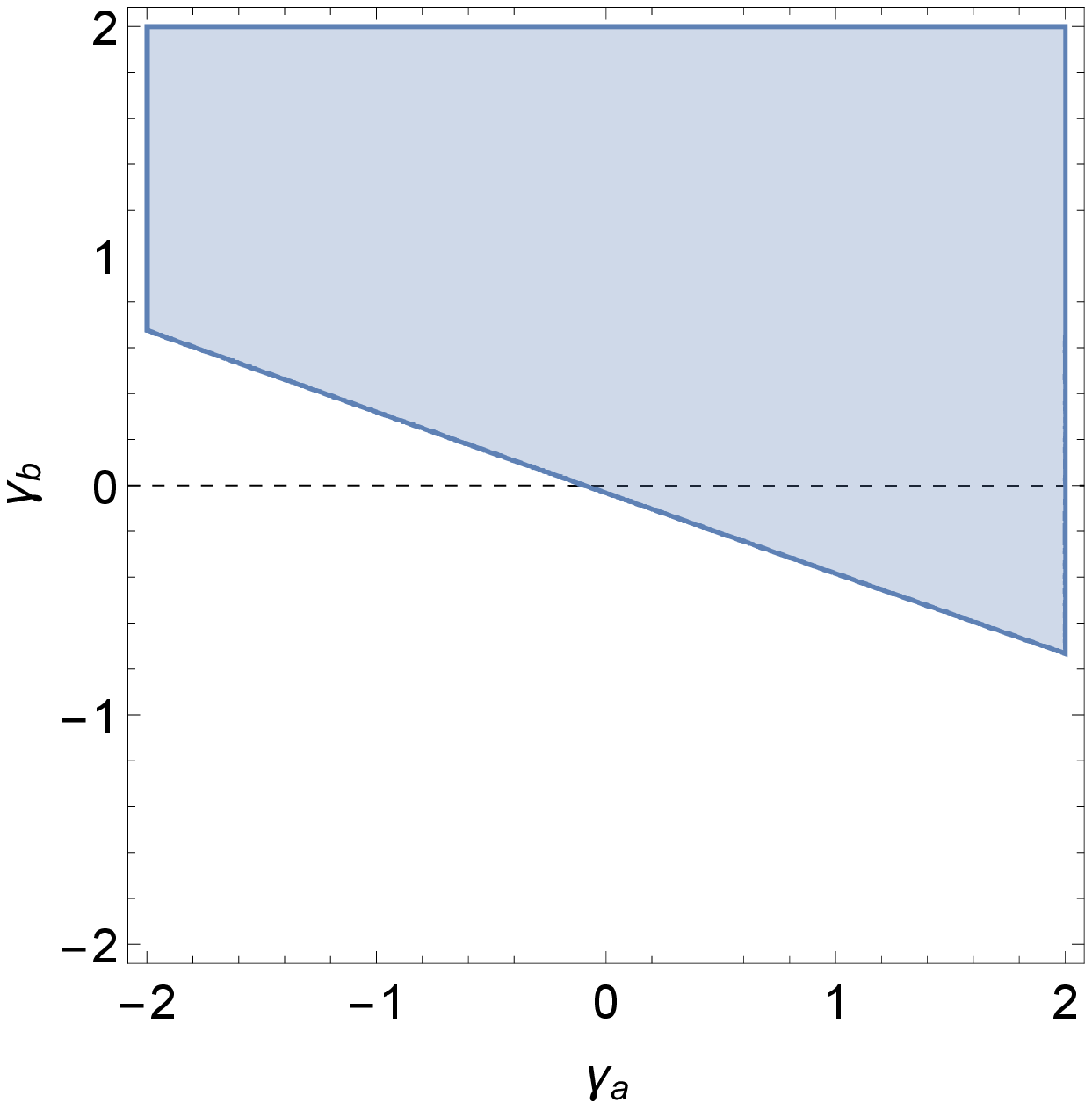}  
    \includegraphics[width=0.32\textwidth]{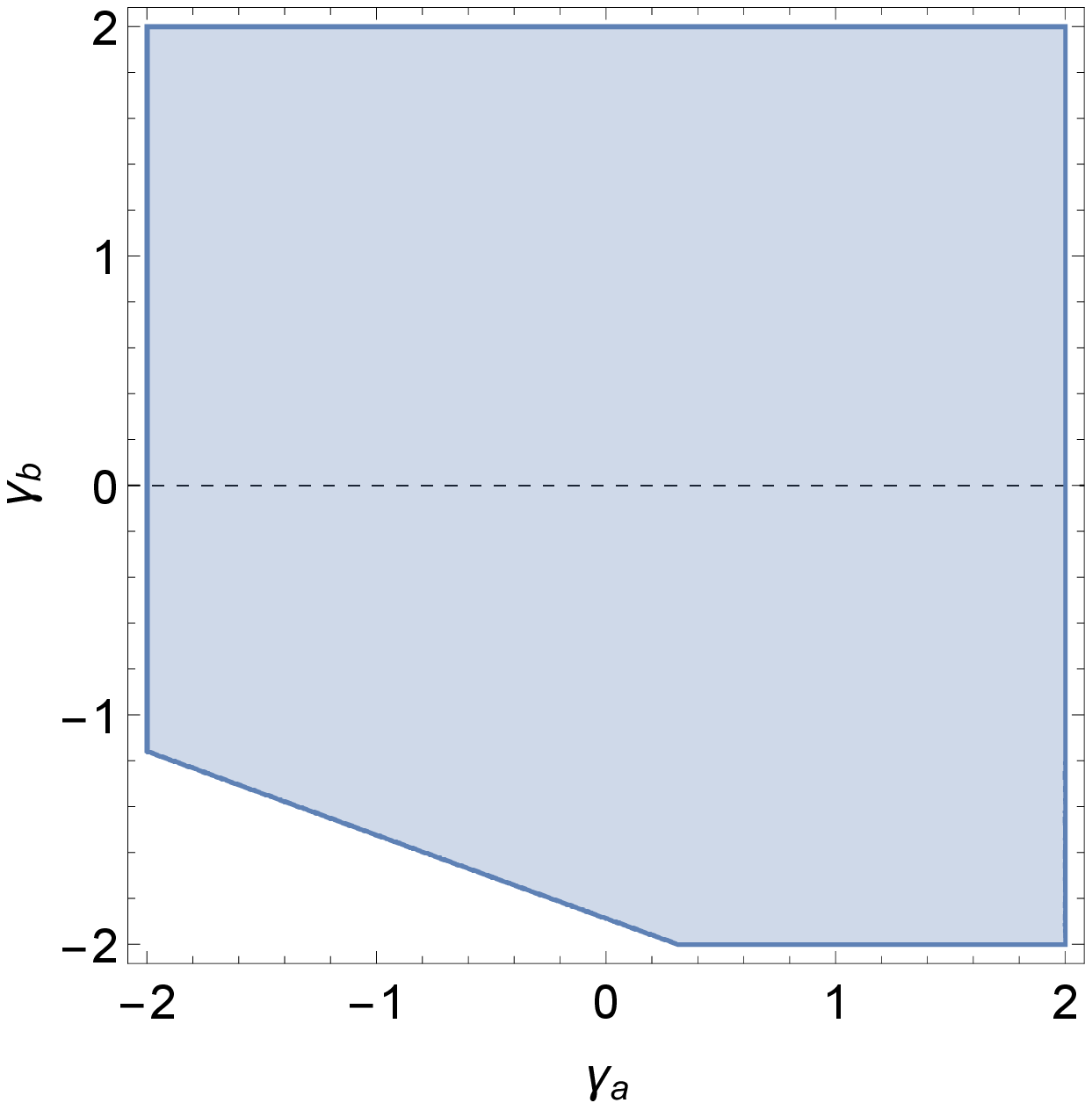}  
\caption{Possible parameter space between $\gamma_a-\gamma_b$ for fixed $a_{\ast}=1$, $L=1$, $k=1$, $\lambda=0.1$: Left $\gamma_c=-1$, Middle $\gamma_c=0$, Right $\gamma_c=1$.}
\label{fig:figMpara}
\end{figure}
In general, the bubble can be either expanded or collapsed after rolling down from the effective potential. However, we are interested only the expanding bubble case which can change the values of the effective CC from the false vacuum AdS space to the true vacuum dS space asymptotically. Therefore, a consideration of the collapsing bubble case is neglected in this study. One can see Ref.\cite{Camanho:2013uda} for more detail discussions of the collapsing bubble that leads to violation of the cosmic censorship hypothesis. 

We close this section by providing a brief discussion on how we ensure that the bubble will expand to infinity reaching the cosmic horizon and changing the effective CC to the new value a whole spacetime. By considering the definitions of the canonical momenta of the AdS (outside) and dS (inside) spaces and the junction condition in (\ref{eqg}) and (\ref{junction-con}), respectively, we find,
\begin{eqnarray}
H \approx \frac{1}{2\left( \mathcal{M}_+ - \mathcal{M}_-\right)}\left[ a^4\int_{g_-}^{g_+}dg\,\Upsilon\big[g\big] - \Big[g\left(\mathcal{M} -a^3\gamma_a -a^2\gamma_b -a\gamma_c \right)\Big]_-^+\right].
\end{eqnarray}
Here the above expression is obtained up to the first order expansion of the  $1/H$. We can see explicitly that $H\equiv \dot{a}/a$ (speed of the bubble expansion) approaches to $\infty$ when $a \to \infty$. This corresponds to the the bubble expansion to infinity at the finite time.  

\section{Gravitational phase transition}\label{sec4}
This section presents a central result of this work on the AdS to dS phase transition mediated by thermalon in EGBMG theory. This section is divided into two parts. First, we discuss how we obtain physically corrected thermalon configuration. Second, from a set of parameter in EGBMG which allowed thermalon configuration, we analyse thermodynamic phase space of the gravitational phase transition.


\subsection{Thermalon configurations, horizons and Nariai bound}
\begin{figure}[h]
    \centering
    \includegraphics[width=0.8\textwidth]{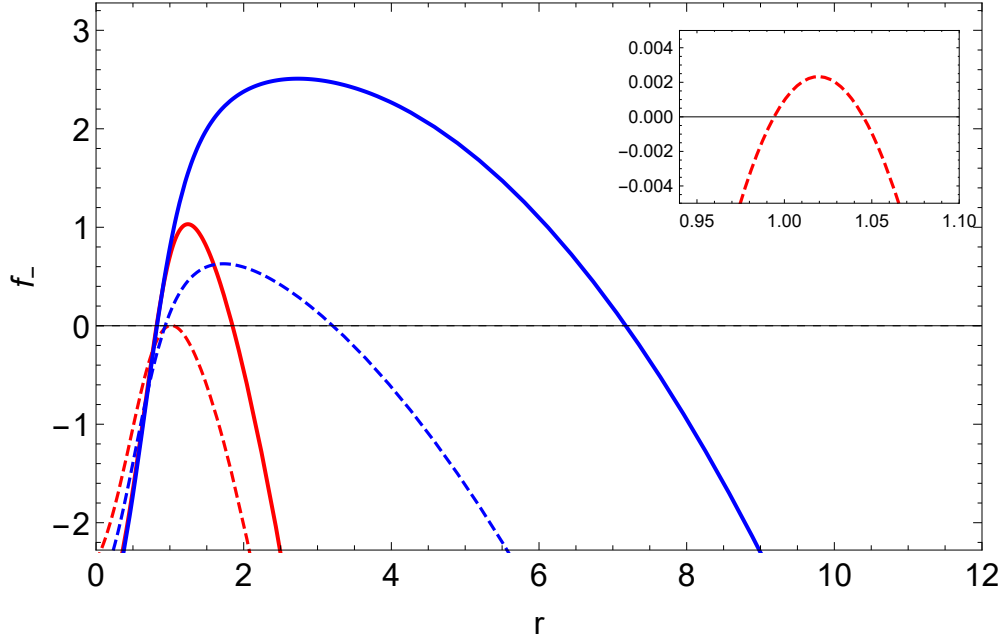}  
\caption{Plot of $f_-$ when $a_{\ast}=1,k=1$ Red: $L=1,\lambda=0.2,\gamma_c=1$ (dashed) $\gamma_a=0,\gamma_b=0$ (solid) $\gamma_a=0.6,\gamma_b=0.5$ and Blue: $L=3,\lambda=0.015,\gamma_c=0.5$ (dashed) $\gamma_a=0,\gamma_b=0$ (solid) $\gamma_a=1.6,\gamma_b=0.8$. The subplot displays a close up behaviour of metric function $f$.} \label{fig:metricfn}
\end{figure}

Since we need to ensure that the thermalon's location $a_{\ast}$ lies in the casual connected region outside the black hole. Thus, it is convenient to firstly determine the casual structure of the metric Eq.(\ref{metricfn}). The horizon's location can be obtained by imposing $f_-(r_{H})=0$. This yields quartic equation
\begin{align}
    r_{H}^4 - \frac{\gamma_a L^2}{3} r_{H}^3 - \left(1+\gamma_b\right)L^2 r_{H}^2 -2\gamma_c L^2 r_{H} + L^2 \left(M^{\ast}_- - \lambda L^2\right) &= 0. \label{quartic}
\end{align}
The roots of this equation can be interpreted as the dS branch (inner spacetime) horizon. In general, there will be four possible roots. However, only real positive roots are physically relevant. An example plot of metric function $f$ is illustrated in fig.~\ref{fig:metricfn}. In these plots, the bubble position is always fixed to unity i.e., $a_{\ast}=1$. Two roots of $f_-$ correspond to event horizon ($r_H$) and cosmological horizon ($r_C$). We observe that as $\gamma_a,\gamma_b$ increases the metric function also increases. In addition, increasing in $\gamma_a,\gamma_b$ does not significantly alter the location of the event horizon while the cosmological horizon is shifted to a larger value. Moreover as we expect, the radius of cosmological horizon becomes larger as $L$ is increased.

Equation (\ref{quartic}) is in a standard form of quartic equation $a r^4 + b r^3 + c r^2 + d r + e =0.$ The roots of this equation are given by

\begin{align}
    r_{1,2} &= -\frac{b}{4a} - S \pm \frac{1}{2}\sqrt{-4S^2-2p+\frac{q}{S}}, \\
    r_{3,4} &= -\frac{b}{4a} + S \pm \frac{1}{2}\sqrt{-4S^2-2p-\frac{q}{S}},
\end{align}
where we have defined the following parameters
\begin{align}
    p &\equiv \frac{8ac-3b^2}{8a^2} = -\frac{L^2}{24}\left(L^2\gamma_a^2+24\left(1+\gamma_b\right)\right), \nonumber \\
    q &\equiv \frac{b^3-4abc+8a^2d}{8a^3} = -\frac{L^2}{216}\left(L^4\gamma_a^3 + 36L^2\gamma_a\left(1+\gamma_b\right) + 432\gamma_c\right), \nonumber \\
    S &\equiv \frac{1}{2}\sqrt{\frac{1}{3a}\left(P+\frac{\delta_0}{P}\right)-\frac{2p}{3}}, \nonumber \\
    P &\equiv \left(\frac{\delta_1 + \sqrt{\delta_1^2 - 4 \delta_0^3}}{2}\right)^{1/3}, \nonumber \\
    \delta_0 &\equiv c^2 - 3 b d + 12 a e = 12\mathcal{M}^*_-L^2 + \left(\left(1+\gamma_b\right)^2 - 2\gamma_a\gamma_c - 12 \lambda\right)L^4, \nonumber \\
    \delta_1 &\equiv 2c^3 - 9 b c d + 27 b^2 e + 27 a d^2 - 72 a c e = \left(72 \mathcal{M}^*_-\left(1+\gamma_b\right) + 108\gamma_c^2\right)L^4  \nonumber \\
    &+ \left( 3\mathcal{M}^*_- \gamma_a^2 - 2 \left(1+\gamma_b\right)\left( 1 + \gamma_b\left(2+\gamma_b\right) -3 \gamma_a\gamma_c + 36\lambda  \right)   \right)L^6 -3\lambda \gamma_a^2 L^8.
\end{align}

\begin{figure}[h]
    \includegraphics[width=0.32\textwidth]{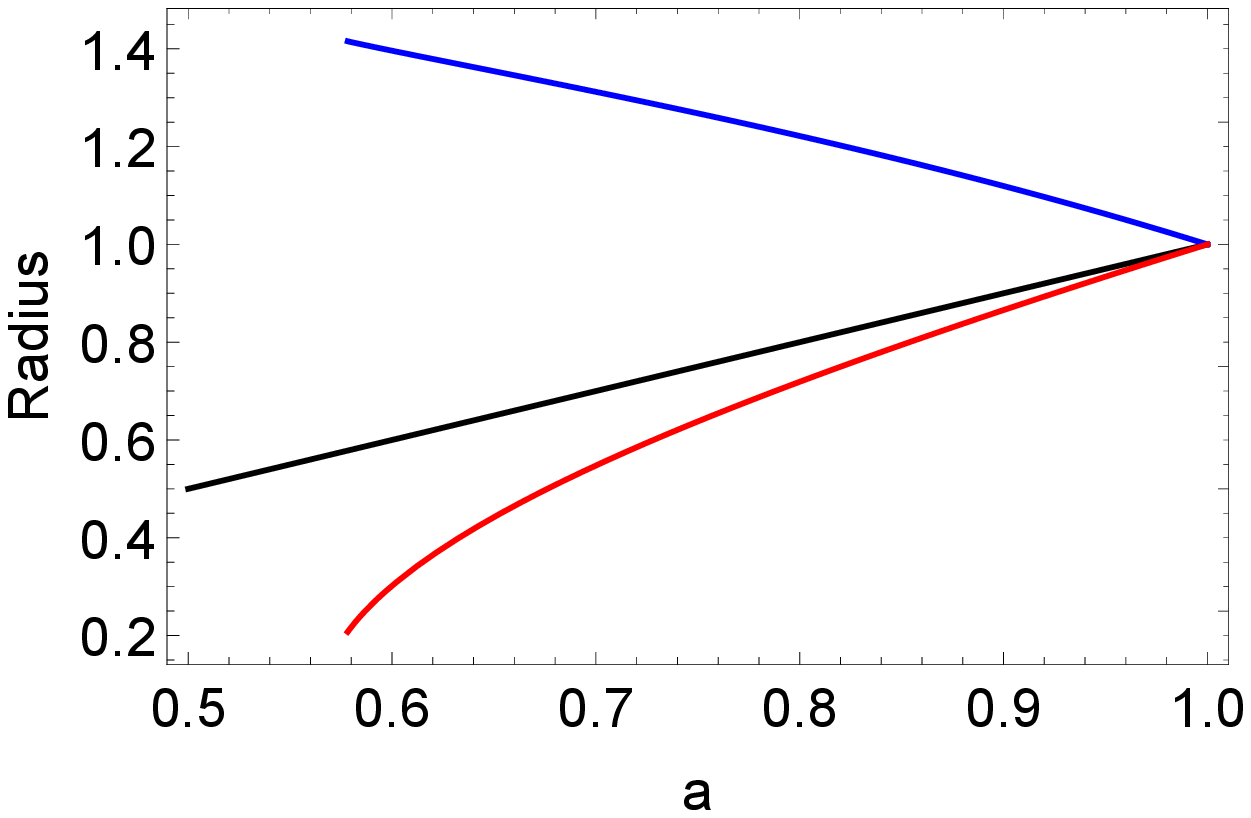}  
    \includegraphics[width=0.32\textwidth]{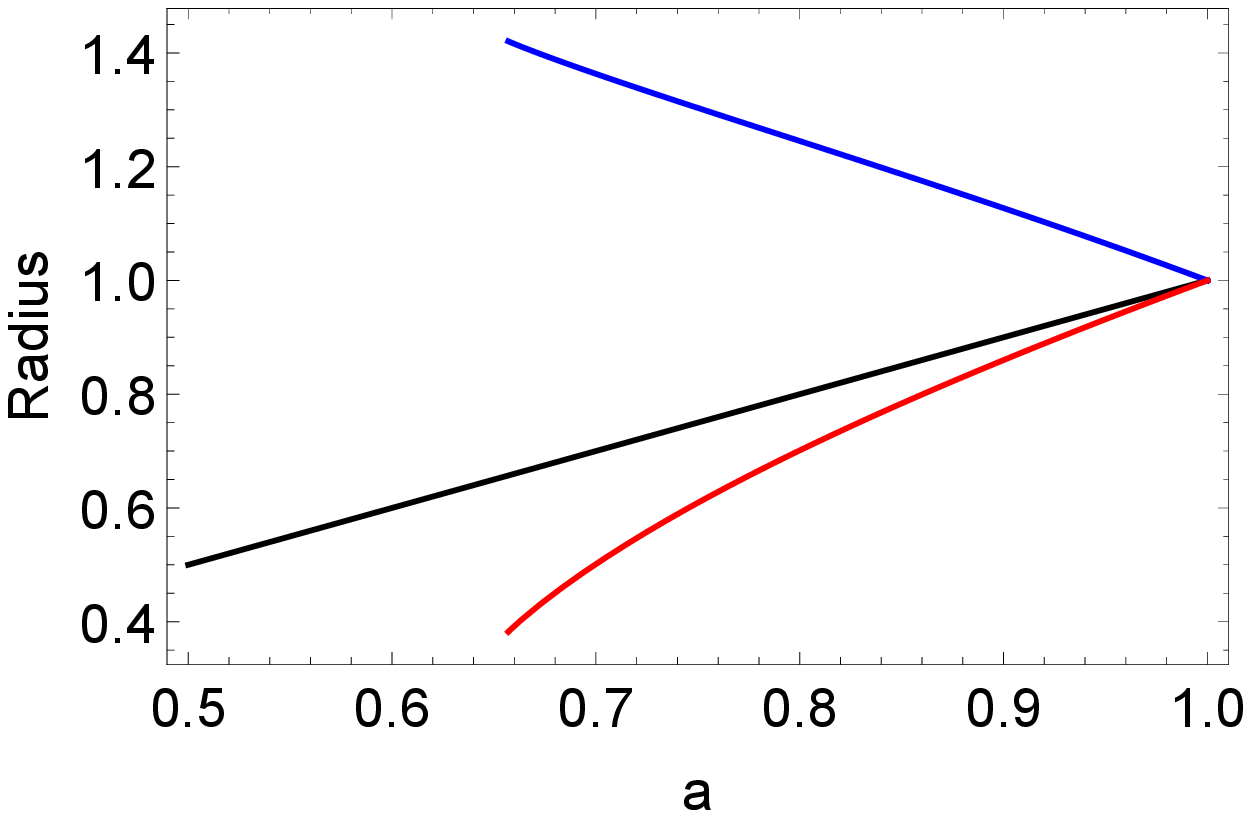}  
    \includegraphics[width=0.32\textwidth]{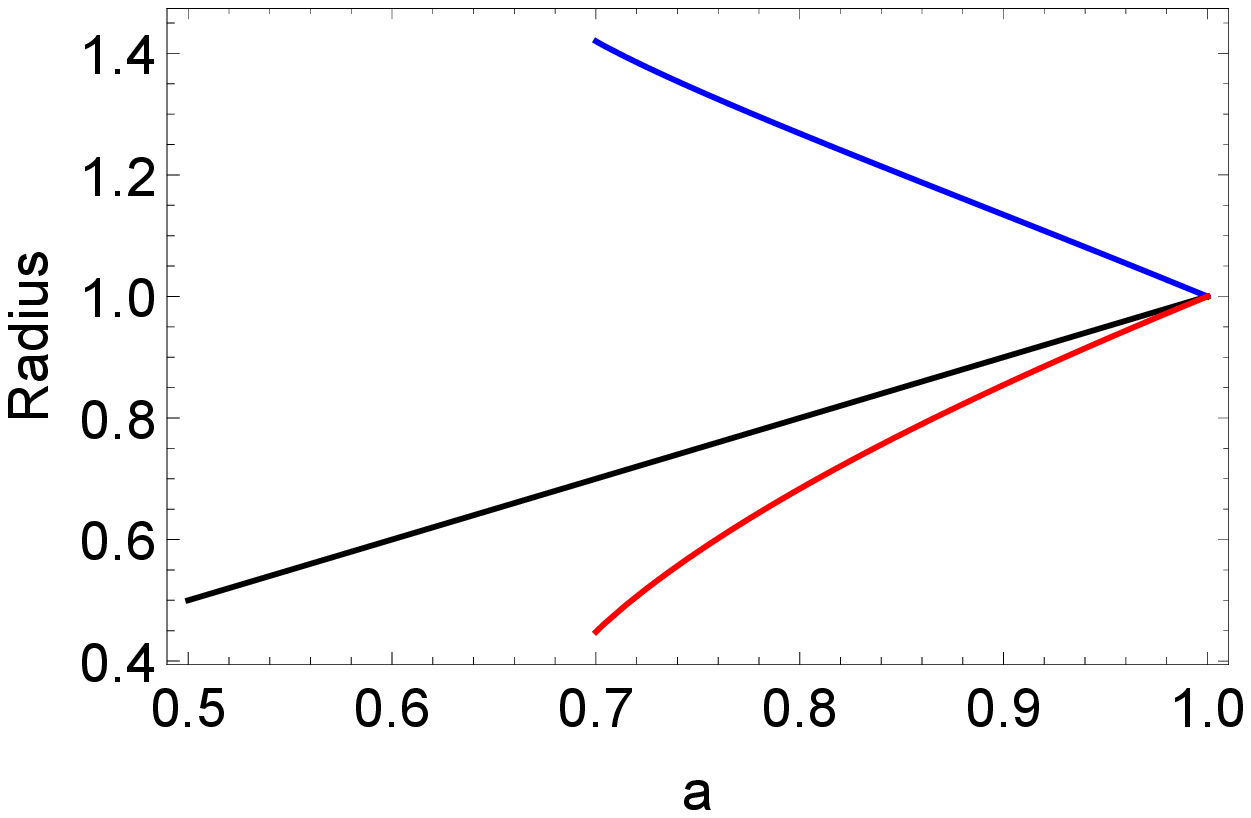}  
\caption{Location of horizons as a function of $a_{\ast}$ when $k=1$, $L=1$,  $\lambda=0.2$, Left: $\gamma_a=-2.393,\gamma_b=2.197,\gamma_c=0$, Middle: 
$\gamma_a=-0.931$, $\gamma_b=-0.035$, $\gamma_c=1.5$, Right: $\gamma_a=0.532$, $\gamma_b=-2.266$, $\gamma_c=3$. The red, black and blue curved are event horizon, thermalon (bubble) location and cosmological horizon respectively. } \label{fig:fig5}
\end{figure}

Thus we can now plot the horizon radius as a function of thermalon's location $a_\ast$ and this is demonstrated in fig.~\ref{fig:fig5}. To be physically relevant, we pay our attention only to the case where the thermalon is located within the casual region of inner spacetime i.e. $r_H \leq a_\ast \leq r_C$. In practice, we fix the maximum value of bubble location at $a_\ast = 1$ where the spacetime approaches the Narai limit $r_H \sim r_C$. We remark that, throughout this work, we shall especially focus on the case where 1) both $\mathcal{M}^*_{\pm}$ are real positive. 2) the metric function Eq.(\ref{metricfn}) has at least two positive real roots. 3) The thermalon always stay between black hole's event horizon and cosmological horizon. We need to ensure that these three conditions are met throughout our investigation. 

\subsection{Thermodynamics of the AdS to dS phase transition mediated by thermalon}
This subsection aims to investigate the gravitational AdS to dS phase transition in the thermodynamics phase space. We follow Refs.\cite{Camanho:2015zqa,Camanho:2013uda,Camanho:2012da,Hennigar:2015mco,Samart:2020mnn,Samart:2020qya} for all formalisms in this study. The gravitational phase transition in this work is a generalization of the well known Hawking-Page (HP) phase transition \cite{Hawking:1982dh} where at the critical temperature the thermal AdS space becomes unstable and then decays to the the AdS black hole with the lower free energy. We employ the generalized HP phase transition in this work. The decay mechanism proceeds via nucleation of the bubbles or the thermalon mediation where the initial false vacuum state is the thermal AdS. When it is unstable and decay into black hole inside the bubble of true vacuum dS space at the critical temperature. This phase transition is produced by the quantum tunneling or jumping over the wall of the quasi particle state in the Euclidean sector of the bubbles called the thermalon \cite{Gomberoff:2003zh,Camanho:2012da,Camanho:2013uda}. In other words, the thermalon effectively changes the branch of the false vacuum AdS solution to another true vacuum dS solution.  

Taking this phase transition mechanism into account, one can calculate the thermodynamics properties of the thermalon in the same fashion as black hole thermodynamics are calculated. The Euclidean action of the thermalon, $\mathcal{I}_E$ has been derived in \cite{Camanho:2012da,Camanho:2013uda} and it is given by 
\begin{eqnarray}
\mathcal{I}_E = \beta_+\,\mathcal{M}_+ + S\,,
\end{eqnarray}
where $\beta_+$, $\mathcal{M}_+$ and $S$ are the inverse Hawking temperature, mass of the external observer in the asymptotic thermal AdS and the entropy of the dS black hole, respectively. The Gibbs free energy of the thermalon in the canonical ensemble is written by  \cite{Camanho:2015zqa,Camanho:2013uda,Frassino:2014pha},
\begin{align}
    F &= \mathcal{M}_+ - T_+ S.
    \label{free-energy}
\end{align}
To obtain the Gibbs free energy of the thermalon, one can use the on-shell regularization method by subtracting the thermal AdS space background \cite{Camanho:2015zqa,Camanho:2013uda,Hennigar:2015mco}. More importantly, there are two competing phases in this system i.e., the thermal AdS (outside) and the thermalon hosting the black hole in the dS space (inside). Then, we will study the phase transition by comparing the Gibbs free energy of the thermalon to the thermal AdS one. In practice, we can set the Gibbs free energy of the thermal AdS space as zero ($F_{\rm AdS} = 0$) since $F_{\rm AdS}$ has already considered as the background subtraction.
The Hawking temperature of inner five dimensional dS black hole in EGBMG can be defined by \cite{Cai:2003kt}
\begin{align}
    T_- &= \frac{1}{4\pi r_H}\left[ \sum_{n=0}^{2} \frac{ k c_n \left(4-2n\right)}{\left(k r_H^{-2}\right)^{1-n}}  + \gamma_a r_H + 2 \gamma_b + \frac{2\gamma_c}{r_H}  \right]\left[\sum_{n=0}^2 \frac{nc_n}{\left(kr_H^{-2}\right)^{1-n}}\right]^{-1}, 
    \label{T-}
\end{align}
where
\begin{align}
    c_0 &= -L^{-2},~~~~~c_1 = 1,~~~~~c_2 =\lambda L^2. \nonumber
\end{align}
Moreover, the Bekenstein-Hawking entropy is given by  \cite{Cai:2003kt}
\begin{align}
        S &= 4\pi r_H^5 \sum_{n=0}^2 \frac{nc_n\left(kr_H^{-2}\right)^n}{k\left(5-2n\right)}.
        \label{entropy}
\end{align}
The Hawking temperature of outer AdS spacetime is computed by
\begin{align}
    T_+ &= \frac{\sqrt{f_+(a_\ast)}}{\sqrt{f_-(a_\ast)}}T_-.
    \label{T+}
\end{align}
\begin{figure}[h]
    \includegraphics[width=0.45\textwidth]{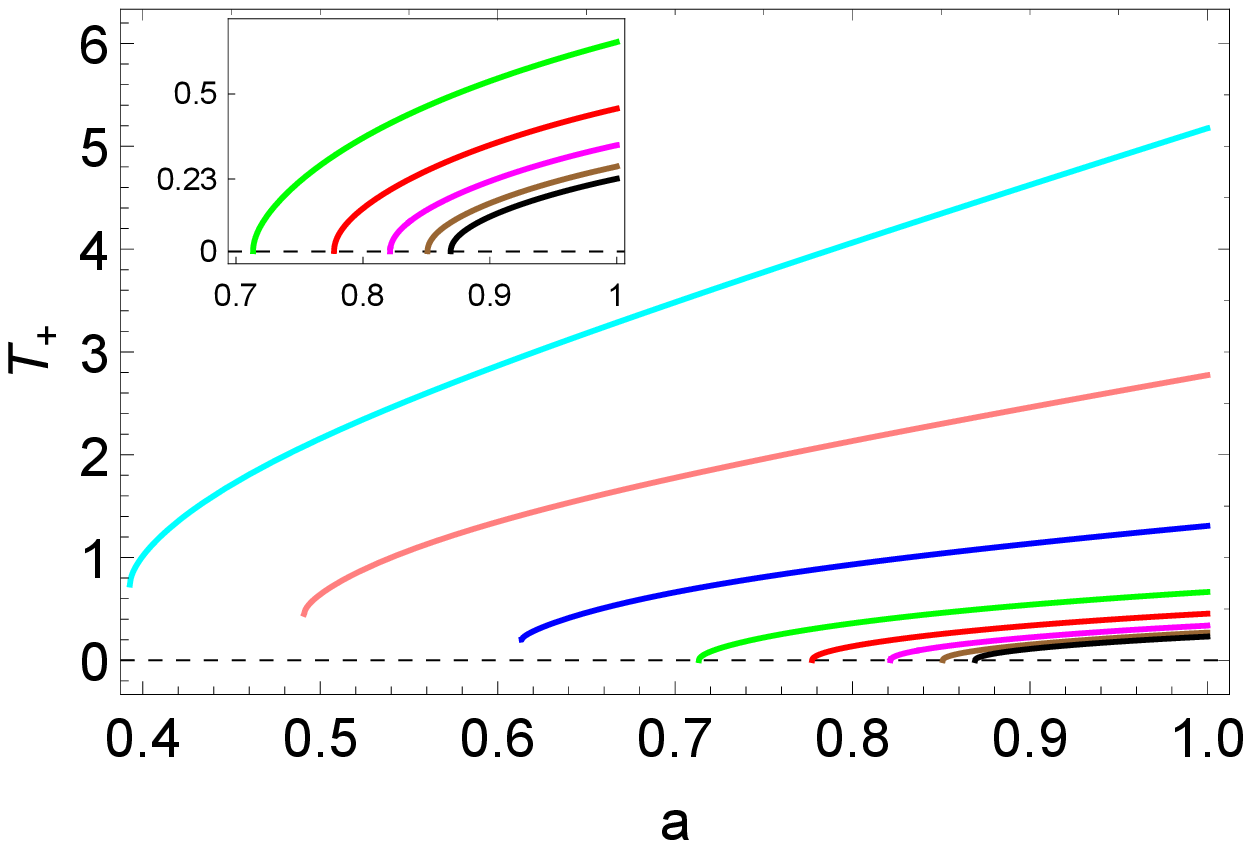}  
    \includegraphics[width=0.47\textwidth]{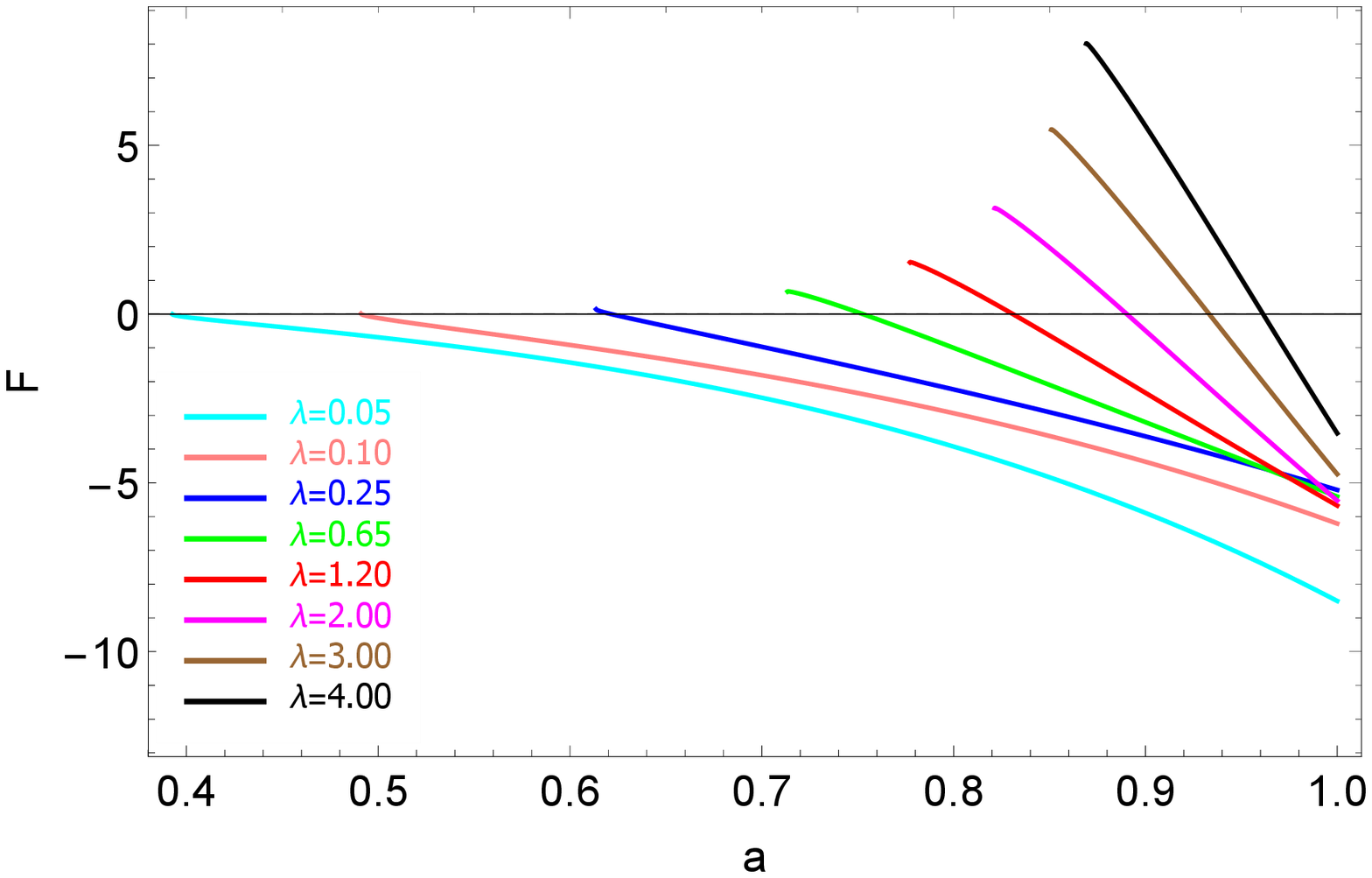}  
\caption{The Hawking temperature (left) and free energy (right) of thermalon configuration as function of $a$ for various value of $\lambda$ with $k=1$, $L=1$, $\gamma_a=-2.296$, $\gamma_b=2.048$ and $\gamma_c=0.1$.} \label{fig:fig6}
\end{figure} 
\begin{figure}[h]
    \includegraphics[width=0.8\textwidth]{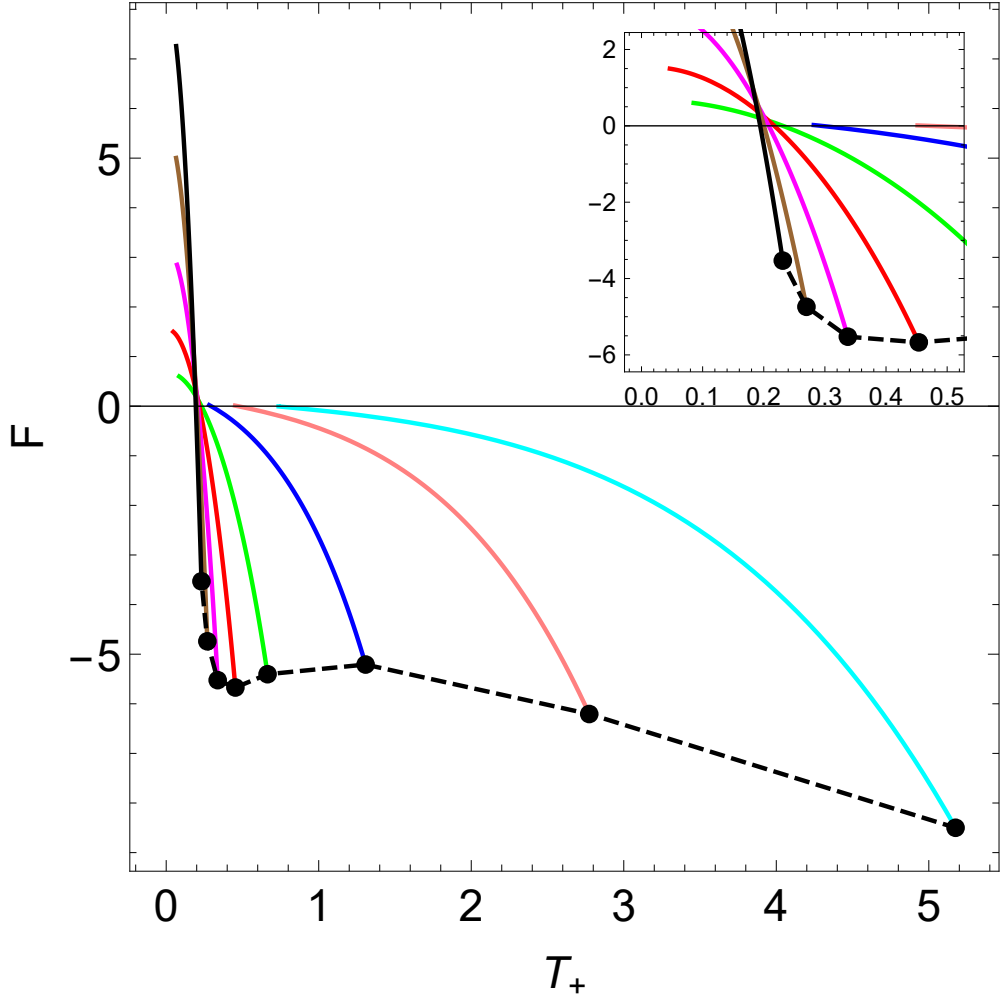}  
\caption{Free energy $F$ of the thermalon configuration as a function of the themperature $T_+$ for various value of $\lambda$ with $k=1$, $L=1$, $\gamma_a=-2.296$, $\gamma_b=2.048$ and $\gamma_c=0.1$. The subplot displays the behaviour of free energy at larger value of $\lambda$. The colored code can be found in Fig.~\ref{fig:fig6} } \label{fig:fig7}
\end{figure} 
We plot $T_+$ and $F$ as a function of the thermalon radius, $a$ and the results are shown in Fig.~\ref{fig:fig6}. In Fig.~\ref{fig:fig6}, the temperature and free energy are plotted as a function of the thermalon's location. It is clear that the temperature increases with bubble position. As $\lambda$ increases, the slope of $T_+$ becomes more flatten. Moreover, we observe phase transition in free energy plots. As $a$ increases, the thermalon configuration becomes more thermodynamically preferred. The phase transition is apparently more obvious as $\lambda$ is larger. The phase transition occurs at larger bubble's location as $\lambda$ increases. Additionally, we observe that the physically relevant thermalon configuration exists at a smaller thermalon's location $a$ at lower $\lambda$. 

By substituting results in Eqs.(\ref{thermalon-mass},\ref{T-},\ref{T+},\ref{entropy}) in the thermalon free energy Eq.(\ref{free-energy}), we are ready to study the thermodynamics phase space between $F$ as function of $T_+$.  
The relation between the Gibbs free energy and Hawking temperature $T_+$ is illustrated in Fig.~\ref{fig:fig7}. In this figure, the parameter $\lambda$ is varied while the others are kept fixed. As $\lambda$ increases, the free energy become less dependent on temperature (the slope becomes more inclined). According to our numerical investigation, we particularly choose the value of $\gamma_a=-2.296, \gamma_b=2.048$ and $\gamma_c=0.1$ such that the thermalon's location lies between the cosmological and event horizons. At small $\lambda$, the minimum value of $F$ is increasing as $\lambda$ becomes larger. However, the minimum starts to decrease from $\lambda=0.25$ until it reaches the lowest value at $\lambda=1.2$. After that, the minimum of free energy rapidly increases as $\lambda$ increases. 

As a result, we find that thermalon always produces the AdS to dS phase transition since all Gibbs free energies of the thermalon are negative at their maximum temperature for all possible values of the given Gauss-Bonnet coupling, $\lambda$. This is an interesting result in this work because usually, the Gauss-Bonnet coupling, $\lambda$ plays the role as the order parameter of the AdS to dS phase transition in both without \cite{Camanho:2015zqa,Hennigar:2015mco} and with the presences of the gauge fields \cite{Samart:2020qya,Samart:2020mnn}. Our results suggest that with a proper value of $\gamma_{a,b,c}$, the gravitational phase transition from AdS to dS is inevitable. This means that in this scenario, the couplings of the massive gravity, ($\gamma_{a,b,c}$) play the role as the order parameter rather than the Gauss-Bonnet coupling, ($\lambda$) in this type of phase transition.

\section{Conclusion}\label{sec5}

In recent years, the most widely studied theory in the higher curvature gravities is known as Einstein-Gauss-Bonnet (EGB) theory of gravity. The EGB theory of gravity is a special case of the Lovelocks gravity. The GB term is only nontrivial in spacetime dimensions $(4+1)D$ or greater. In this work, we have studied gravitational phase transition from AdS to dS geometries by engaging the well-defined massive theory of gravity to the EGB one. This inclusion basically features a matter field sector in the theory.

The thermalon induces the decay of the negative effective cosmological constant to the positive one due to the higher-order gravity effects and the decay mechanism proceeds through nucleation of the bubbles or the thermalon of true vacuum (dS) inside the false vacuum (thermal AdS) Refs.\cite{Camanho:2012da,Samart:2020qya}. It was demonstrated that the probability of the decay, $P$, of
the thermalon effectively jumping from AdS to dS branch
solutions is governed by $P\propto e^{I_{E}}$ with $I_E$ being the Euclidean action difference between initial thermal AdS and the thermalon (bubble state). As a result, the system will end up in the stable dS black hole after the thermalon
expansion reaching the asymptotically dS region in a
finite time.

We have also determined the thermodynamics phase space of the Hawking temperature and free energy and found that the free energy of the thermalon is always negative at the maximum of Hawking temperature for all specific values of the parameters of the massive gravity sector. Additionally, we have also discovered that a generic behavior of the phase transition in the EGBMG theory is affected by the specific values of the parameters in the massive gravity instead of the GB coupling in contrary to those cases found in the previous publications in which the authors have studied the vacuum and the charged cases.  

With presence of the massive gravity sector, we have demonstrated that the phase transition mediated by thermalon from AdS to dS asymptotics is inevitable in the present work. and specific valued of the parameters in the massive gravity behave as the order parameters in the phase transition instead of the Gauss-Bonnet coupling. Interestingly, our results are in line with the claim that the generalized gravitational phase transition is a generic behavior of the higher-order gravity theories when the matter field is included.

\begin{acknowledgments}
This work (Grant No. RGNS 64-217) was financially supported by Office of the Permanent Secretary, Ministry of Higher Education, Science, Research and Innovation. We are graceful to Chakrit Pongkitivanichkul for useful discussion. P. Channuie acknowledged the Mid-Career Research Grant 2020 from National Research Council of Thailand (NRCT5-RSA63019-03). D. Samart is financially supported by the Mid-Career Research Grant 2021 from National Research Council of Thailand under a contract No. N41A640145. S.P. would like to acknowledge Piyabut Burikham for his advice and support. 
\end{acknowledgments}

\bibliography{GBphase}


\end{document}